\documentclass[prl,aps,twocolumn,showpacs]{revtex4}

\usepackage[dvips]{graphicx}
\usepackage{dcolumn}%
\usepackage{amsmath}%
\usepackage{amsfonts}%
\usepackage{amssymb}
 \usepackage[usenames]{color}
\usepackage{appendix}

\providecommand{\U}[1]{\protect\rule{.1in}{.1in}}
\newcommand{\be}{\begin{equation}}
\newcommand{\ee}{\end{equation}}
\newcommand{\bea}{\begin{eqnarray}}
\newcommand{\eea}{\end{eqnarray}}
\newcommand{\bt} {\begin{tabular}}
\newcommand{\et} {\end{tabular}}
\newcommand{\nn}{ \nonumber}
\newcommand{\ds}{\displaystyle}
\newcommand{\ba} {\begin{array}}
\newcommand{\ea} {\end{array}}
\begin{document}

\title{The effect of Coulomb interactions \\of thermoelectric characteristics  of Marcus molecular junctions}

\author{  Natalya A. Zimbovskaya$^{1,2}$}

\affiliation
{$^1$Department of Physics and Electronics, University of Puerto Rico,  Humacao, PR 00791, USA and $^2$Department of Chemistry, University of Pennsylvania, PA 19104, USA}

\begin{abstract}
We present a theoretical study focused on the influence of Coulomb interactions on thermoelectric properties of molecular junctions assuming that electron transport is strongly affected by thermalized phonon modes in the molecular ambient. It is shown that the combined effect of Coulomb interactions between  electrons on the molecular states and the reorganization processes in the ambient may significantly affect both Seebeck coefficient and the power factor. Specifically, we show that electron-electron interactions may counter-balance the effect of the ambient molecules reorganizations resulting in qualitative changes in the behavior of both zero-bias conductance and thermopower.
\end{abstract}

\date{\today}
\maketitle

   \section
{I. Introduction:} 

The current interest to thermoelectricity in small (nanoscale) systems was triggered by pioneering works \cite{1,2} predicting a dramatic enhancement of the efficiency of heat-to-electricity conversion in these systems. Now, studies of thermoelectric properties of tailored nanoscale systems such as carbon-based nanostructures, quantum dots and metal-molecule junctions represent a well established research field \cite{3,4,5,6,7,8,9} providing a general platform  for development of molecular electronics. Here, we focus on the thermoelectricity in single molecular junctions (SMJ) which could serve as basing building blocks for diverse nanoscale devices.

A single molecular junction consists of a couple of conducting electrodes bridged by a molecule or a single/multiple quantum dot. Thermoelectric electron transport along the bridge is simultaneously driven by electric forces and thermal gradients. Thermoelectric properties of SMGs are controlled by several well established factors including the geometry of the bridge and the specifics of its coupling to the electrodes \cite{10,11,12,13,14}, quantum interference effects \cite{15,16,17,18,19}, interactions between traveling electrons and the  vibrations of the molecular bridge \cite{20,21,22,23,24,2,26} and/or thermalized phonon modes originating from random motions of molecules in the SMJ ambient \cite{27,28,29} and  Coulomb interactions between electrons \cite{11,14,16,30,31,32,33,34}.

In the majority of existing works the effect of Coulomb interactions on the electron transport through the molecules was analyzed within the ballistic or nearly ballistic transport regime where electron-phonon interactions are weak and may be omitted or the effect of phonons could be treated as a perturbation. In the present study we consider the opposite transport regime assuming that electron- phonon interactions are strong. Specifically, we consider the situation when a SMJ operates being immersed in a dielectric solvent and the electron transport could be visualized as a sequence of hops between the bridge energy levels. Traveling electrons may be temporary localized thus distorting the nearby solvent and giving rise to reorganization processes. Thermalized phonon modes accompanying these processes could strongly interact with the traveling electrons controlling the transport properties of the system.

In the above described situation the electron transport through a SMJ could be theoretically analyzed applying the Marcus theory \cite{35,36,37} which may be modified to account for the lifetime broadening of the bridge levels \cite{38,39}. The effects of solvent reorganization on the charge and heat transport through molecules were (and still are) usually studied using the Marcus theory in its original or modified form (see e.g Refs. \cite{40,41,42,43,44,45,46,47,48,49}). However, in  the most of the works analyzing the transport in Marcus junctions the possible effects of Coulomb interactions are ignored.

      The purpose of the present study is to show that Coulomb interactions may leave their signatures in the characteristics of thermoelectric transport through Marcus SMGs. 
The analysis is carried on assuming that the applied thermal gradient $\Delta T$ is much smaller than the average temperature of the junction $T$. Under this condition the junction response remains linear in $\Delta T$ and both electron conductance $ G $ and thermopower  $ S $ do not depend on temperature. As  known, at greater thermal gradient, the system may switch to the nonlinear regime of operation \cite{16,33,34} but in the present  work we do not extend our studies beyond the linear regime.

\section
{\bf II. Main equations:}
In the following analysis we assume that the bridge is weakly coupled to the electrodes but its coupling to the solvent phonons is strong. The electron-phonon coupling is treated semiclassicaly as it was done in earlier works \cite{35,38,40,48}. Within Marcus approach the steady state electron transport could be analyzed using rate equations  for the probabilities associated with the bridge states. Here, we adopt a simple model postulating  that the bridge may hold no more than two electrons. The schematics of the model is presented in Fig. 1. Then the linker may be simulated by four states, namely, the neutral state $|b>$, two states singly charged by electrons with different spin orientations $|a>_+$ and $|a>_-$ and the doubly charged state $|c>.$

\begin{figure}[t] 
\begin{center}
\includegraphics[width=7.4cm,height=4.2cm]{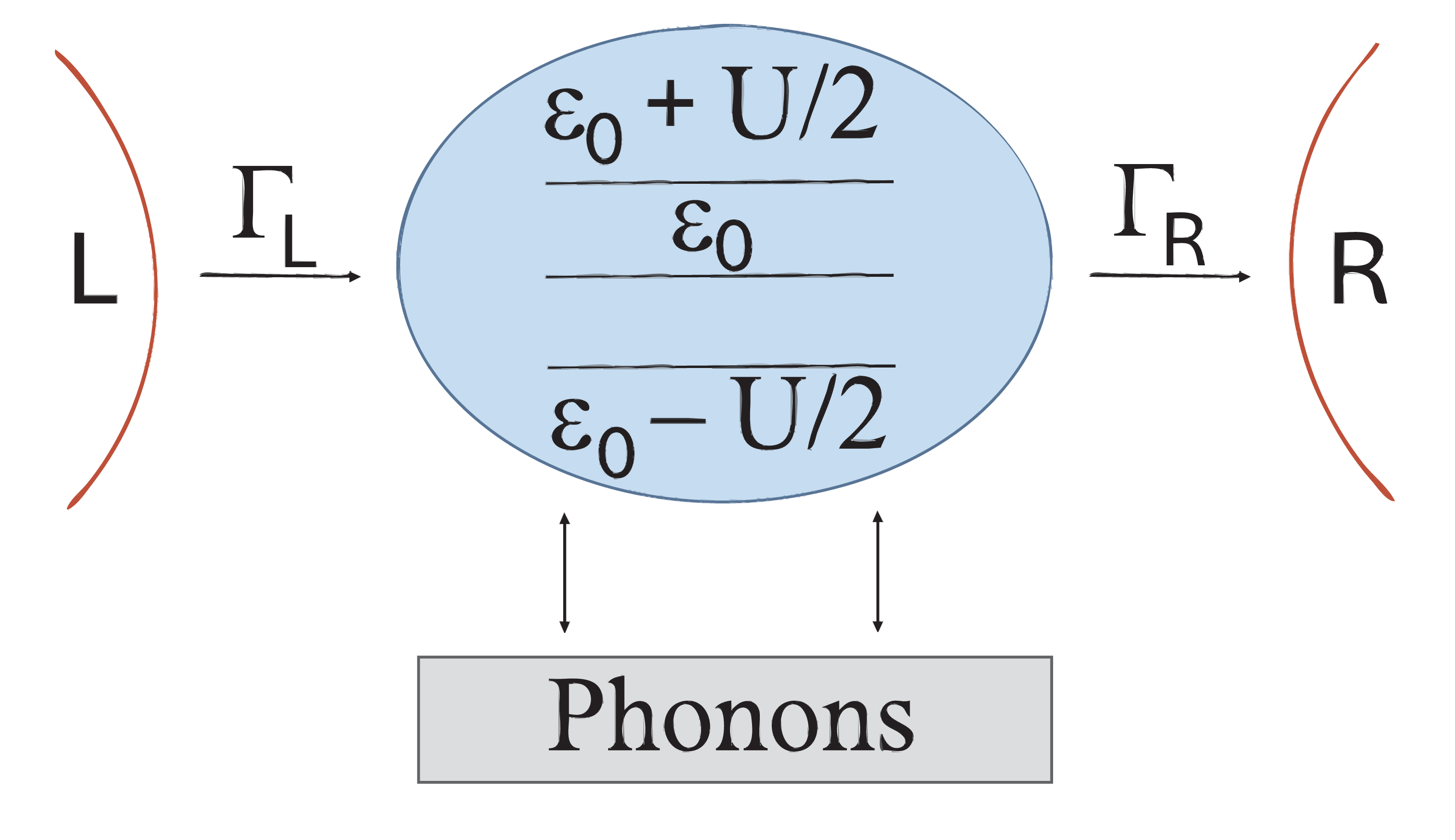}
\caption{Schematics of the adopted model. When there are two electrons on the bridge, the Coulomb repulsion between them splits the bridge energy level with the energy $\epsilon_0$ in two levels with the energies $\epsilon_0\pm U/2$. The linker is coupled to the electrodes with the coupling energies $\Gamma_{L,R}$ The rectangle represents a bath of thermalized phonon modes affecting the electron transport.
}
\label{rateI}
\end{center}
\end{figure}

Note that if the Coulomb interactions are omitted, both singly charged states $|a>_{\pm} $ and the doubly charged state $|c>$ are characterized by the same energy $\epsilon_0 = E_a - E_b \ (E_{a,b}$ are the energies associated with the singly charged and neutral states of the molecule, correspondingly). On the contrary, if the Coulomb interactions are taken into accont, the states $|a>_{\pm} $ are characterized by the energy $\epsilon_a = \epsilon_0 - U/2 $ whereas the energy $ \epsilon_c = \epsilon_0 + U/2$ is attributed to the double charged state $|c>.$ Here, $U $ is the charging energy originating from the Coulomb interactions between electrons on the bridge.
In further  analysis we assume that charging/discharging of electrodes does not affect the solvent. Solvent reorganization solely occurs when electrons are transferred from the electrodes to the molecular bridge and back and is characterized by the reorganization energy $\lambda. $ 

Probabilities for the molecular linker states $ P_a^{\pm}, \ P_b $ and $ P_c $ are determined by rate equations. In writing these equations we should keep in mind that single charged states $|a>_+$ and $|a>_-$ become occupied with equal probabilities $(P_a^{+} =  P_a^{-}) $, therefore we further denote both of them as $ P_a. $ 
     Also, due to the Coulomb interactions the probability $P_c$ may take on nonzero values only provided that $P_a$ differs from zero. Thus the rate equations are:
\be
\frac{dP_a}{dt} = P_b k_{ba} - P_a k_{ab} ; \ \ \  \
\frac{dP_c}{dt} = P_a k_{bc} - P_c k_{cb}   \label{1}
\ee
 where $k_{\alpha b} = k_{\alpha b}^L + k_{\alpha b}^R , \ 
k_{b\alpha } = k_{b\alpha }^L + k_{b\alpha }^R\ \ (\alpha = \{a,c\}), \ \ k_{\alpha b}^K $ and $ k_{b\alpha}^K \ \ (K = \{L,R\} $ denotes the left/right electrode) are the transfer rates associated with the removal/injection of a single electron from/to the molecule. Note that the transfer rates   associated with the states $|a>_+ $ and $|a>_-$ are equal and referred to as $k_{ab}^K$ and $k_{ba}^K$. Using Eqs. (\ref{1}) and the condition $2P_a + P_b + P_c = 1 $ we get the expressions for the steady state probabilities \cite{30,31}:
\begin{align}
 P_b = & \frac{1}{\ds 1 + \frac{2k_{ba}}{k_{ab}}
+ \frac{k_{bc}}{k_{cb}}  \frac{k_{ba}}{k_{ab}}}; 
\nn \\ \nn \\
P_a = & P_b \frac{k_{ba}}{k_{ab}};  \  \  \  \  \
P_c = P_a \frac{k_{bc}}{k_{cb}}  
\label{2} .
\end{align}

The transfer rates $ k_{\alpha b}$ and  $ k_{b \alpha }$ may be presented in the form which takes into account the lifetime broadening of the molecular levels \cite{38,39}:
\begin{align} &
k_{\alpha b}^K = \frac{1}{\pi \hbar} \int_{- \infty}^\infty d \epsilon \Gamma_\alpha^K (\epsilon) \big[1 - f_K(\epsilon) \big] R_-^\alpha (\epsilon);  \label{3}
\\  \nn \\ &
k_{b\alpha}^K = \frac{1}{\pi\hbar} \int_{- \infty}^\infty d\epsilon \Gamma_\alpha^K (\epsilon) f_K (\epsilon)  R_+^\alpha (\epsilon). \label{4}
\end{align}

Here, $\Gamma_\alpha^K (\epsilon)$ describes coupling of the molecular  state $|\alpha> $ to the $ \{L,R\} $ electrodes occurring at the temperatures $ T_K. $ The electrodes are characterized by chemical potentials $\mu_K $ depending on the bias voltage applied across the system. Further we adopt the wide band approximation for the electrodes treating $ \Gamma_\alpha^K $ as constants and assume for simplicity  that the molecular states are symmetrically and equally coupled to the electrodes that is $ \Gamma_\alpha^K \equiv \Gamma.$ Then the functions $R_\pm^\alpha $ accept the form:
\begin{align}
R_{\pm}^{\alpha} = \ &\mathsf{Re}\sqrt{\frac{\pi\beta}{4\lambda}}\exp\Big[\frac{\beta(\Gamma\mp i(\epsilon_{\alpha}\pm\lambda -\epsilon)\big)^2}{4\lambda}\Big] 
\nn\\ & \times
\mathsf{erfc}\Big[\sqrt{\frac{\beta}{4\lambda}}\big(\Gamma\mp i(\epsilon_\alpha \pm\lambda-\epsilon)\big)\Big].   \label{5}
     \end{align}
where $\beta = {1}/{kT}, \ k $ is the Boltzmann constant, $ \alpha = \{a,c\}, $  and $ \mathsf{erfc} (z) $ is the complimentary error function.

Using Eqs. $(\ref {1})$ and $(\ref {2})$ one may write the following expression for the steady state charge current $ I_{ss}: $
\begin{align} &
\frac{I_{ss}}{e} = \frac{\ds 2\Big[I_1 \Big(1 + 
\frac{k_{ba}}{k_{ab}}\Big) + \frac{k_{ba}}{k_{ab}} \Big(1 + \frac{k_{bc}}{k_{cb}}\Big) I_2 \Big]}{\ds1 + 2\frac{k_{ba}}{k_{ab}} + \frac{k_{ba}}{k_{ab}} \frac{k_{bc}}{k_{cb}}}. 
 \label{6}
 \end{align}
 Here:
 \be 
I_1 = \frac{k_{ab}^R k_{ba}^L - k_{ab}^L k_{ba}^R}{k_{ab} + k_{ba}};      \ \ \ \
I_2 = \frac{k_{cb}^R k_{bc}^L - k_{cb}^L k_{bc}^R}{k_{bc} + k_{cb} }.    \label{7}
\ee 
Note that in the limit $U \to 0 $ the considered bridge may be treated as a two level system including a neutral state $|b > $ and a double charged state $|\tilde a > $  which holds a couple of noninteracting electrons with different spin orientations and the occupational probability $ P_{\tilde a} = 2 P_a$.  In this case  $ I_1 = I_2 $ and   Eq.(\ref{6}) is reduced to the form: $I_{ss} = 2eI_1. $

Within the linear transport regime when both bias voltage drop $\Delta V $ and thermal gradient $\Delta T $ applied across a SMJ are small, the charge current may be presented in the form $I_{ss} = G\Delta V + G_{th}\Delta T $ \cite{50} where the electron conductance $\ds G = 
{\partial I_{ss}}/{\partial V} $ and $\ds G_{th} = 
{\partial I_{ss}}/{\partial \Delta T} $ are computed at $\Delta V = \Delta T = 0.$
In this case the thermovoltage $V_{th} $ stopping the charge current is proportional to the thermal gradient, the thermopower $ \ds S = -
{G_{th}}/{G}$ being the coefficient of proportionality independent on $\Delta V$ and $\Delta T$.
     Using Eqs. $(\ref{6}), (\ref{7})$ one way get the following expressions for $ G $ and $ G_{th}: $
\begin{align} 
G = & 2 G_0 \Big [\big(P_b +P_a\big) Q_{ab}^{(0)} + \big(P_a +P_c\big) Q_{cb}^{(0)} \Big];    \label{8}
 \\ \nn  \\ 
 G_{th} = & \frac{2e}{\pi \hbar T} \Big [\big(P_b +P_a\big) Q_{ab}^{(1)} + \big(P_a +P_c\big) Q_{cb}^{(1)} \Big];    \label{9}
 \end{align}
 where
\be
   Q_{\alpha b}^{(n)} = \frac{k_{\alpha b} L_{\alpha,+}^{(n)} + k_{b\alpha} L_{\alpha,-}^{(n)}}{k_{\alpha b} + k_{b\alpha}}
\label{10} \ee
 and
\be
L^{(n)}_{\alpha,\pm}=  -\int_{-\infty}^{\infty}d \epsilon\frac{\partial f}{\partial\epsilon}(\epsilon-\mu)^{n}R_{\pm}^{\alpha}(\epsilon).   
  \label{11}
\ee
In these expressions, $\ds G_0 = \frac{e^2}{\pi\hbar},\, T = T_L = T_R $ and $ \mu = \mu_L = \mu_R $.  

In the limit $U \to 0 $ the expressions for $G$ and $ G_{th} $ are simplified and the thermopower accepts the forms:
\be
S = \frac{1}{T|e|} 
\frac{k_{\tilde ab} L_{\tilde a,+}^{(1)} + k_{b\tilde a} L_{\tilde a, -}^{(1)}}{k_{\tilde a b} L_{\tilde a,+}^{(0)} + k_{b\tilde a} L_{\tilde a,-}^{(0)}} .
\label{12}
\ee
This result coincides with the expression for $S$ given in Ref. 
\cite{39}. Using  Eqs. (\ref{8}) - (\ref{11}) we also may analyze the effect of Coulomb interactions on the power factor $S^2G$  which characterizes the efficiency of heat-to-electricity conversion.

\section
{ III. Results and discussion:}
	First we analyze how Coulomb interactios may affect the behavior of zero bias conduction in Marcus molecular junctions. As shown in Fig. 2a, in the absence of interactions between electrons $ G $ takes on nonzero values over the interval with the end points situated near $ \mu = \epsilon_0 \mp \lambda $   and is represented by a single forked peak with maxima located near these points. The Coulomb interactions  bring the  extension of this interval which in now limited by 
$\ds \mu = \epsilon_0 - 
{U}/{2}  - \lambda$ and $\ds \mu = \epsilon_0 + 
{U}/{2}  + \lambda.$ 
Also, combined effect of Coulomb interactions and reorganization of the solvent molecules increases the number of the conduction maxima. Within the simple model adopted in the present work up to four maxima may appear which  are situated at $\ds \mu= \epsilon_0 - 
{U}/{2}  \mp \lambda$ and $\ds \mu = \epsilon_0 + 
{U}/{2}  \mp \lambda.$ This is illustrated by the lowerest curve displayed in Fig. 2b.

\begin{figure}[t] 
\begin{center}
\includegraphics[width=7.4cm,height=5.5cm]{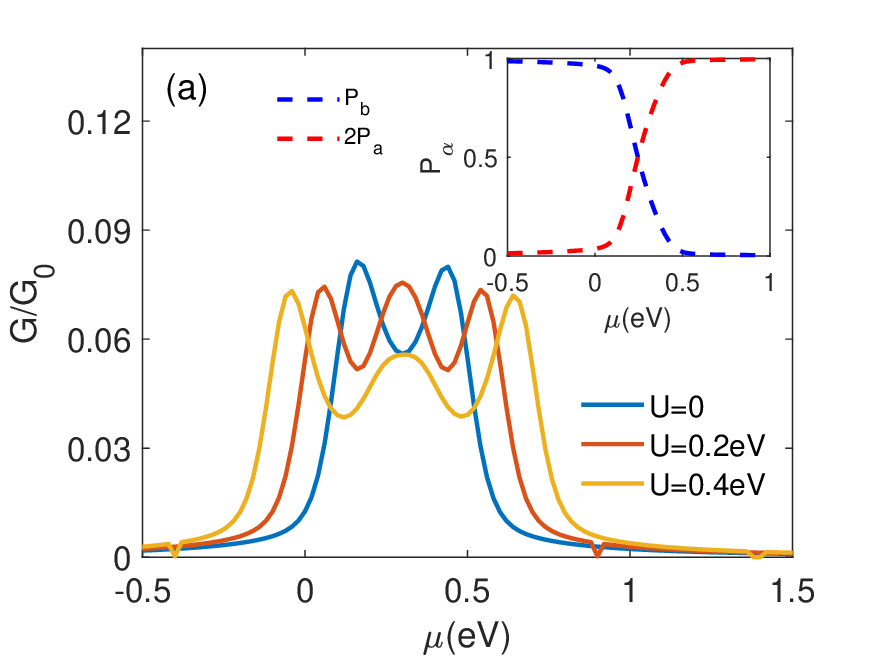}
\includegraphics[width=7.4cm,height=5.5cm]{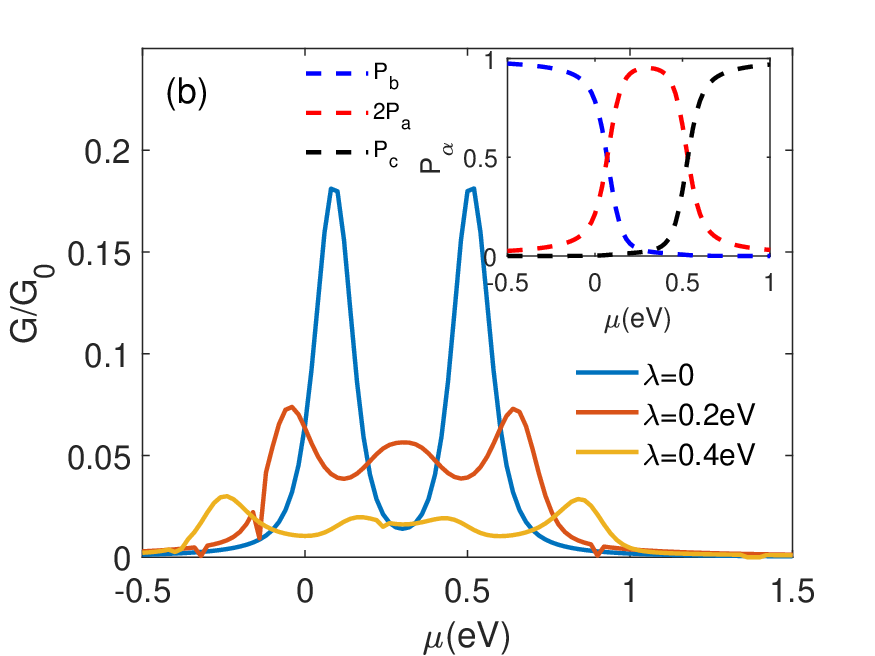}
\includegraphics[width=7.4cm,height=5.5cm]{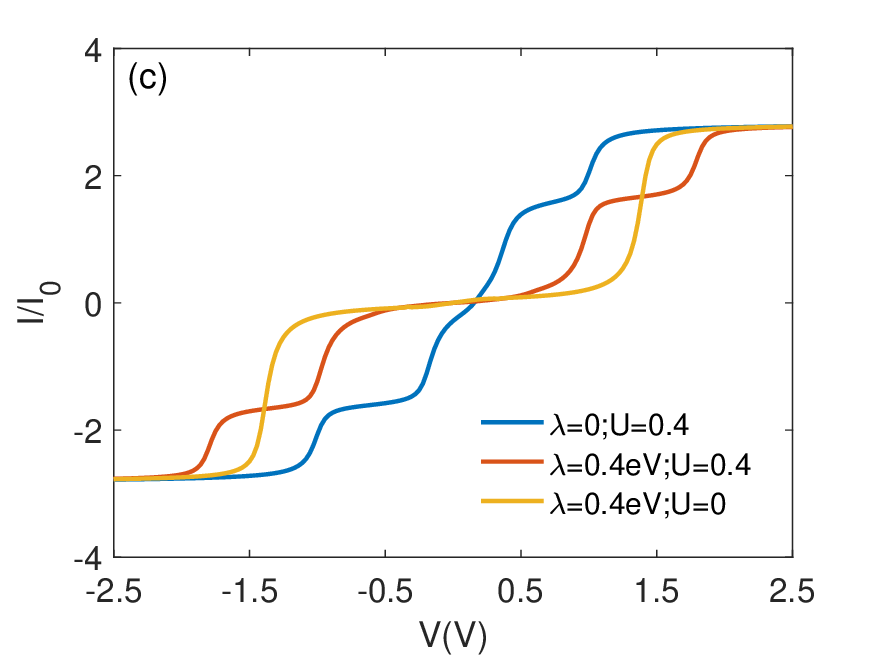}
\caption{Zero bias conductance computed for a Marcus SMJ asuming that $ kT = 0.026 eV,\, \Gamma = \Gamma_L = \Gamma_R = 0.03 eV,\, \epsilon_0 = 0.3 eV,\, \lambda = 0.2 eV $(a) and $U = 0.4 eV $(b). Insets show the occupation probabilities for the bridge states in the limit $U\to  0$(a) at $U = 0.4 eV\,$(b). In the panel (c) the steady state current is presented as a function of the bias voltage $\ds \big(  I_0 = {|e|\Gamma/\hbar}\big).$  The curves are plotted at several values of $ U $ and $ \lambda.$}
\label{rateI}
\end{center}
\end{figure}

To further elucidate how the Coulomb interactions affect the charge transport through the Marcus molecular junction we presented in Fig. 2c the steady state current flowing along the biased SMJ at several values of the chargeing Coulomb constant $U$ and reorganization energy $\lambda$. One observes that Coulomb interactions cause the appearance of the extra step in the $I - V $ curves typical for the electron transport within the Coulomb blocade regime. Also they are  responsible for the decrease of the width of the plateau in the $ I - V $profiles around $V = 0$ which originates from the reorganization of the solvent molecules. Thus one may conclude that the effect of the interactions between the electrons on the bridge opposes the Franck-Condon blocade. In some cases the solvent reaction on the traveling electrons and the Coulomb interactions may counter-balance each other when $ \mu $ approaches $\epsilon_0. $ Then the conductance shows the maximum emerging at $\mu = \epsilon_0, $ as presented in Fig. 2a.

\begin{figure}[t] 
\begin{center}
\includegraphics[width=7.4cm,height=5.7cm]{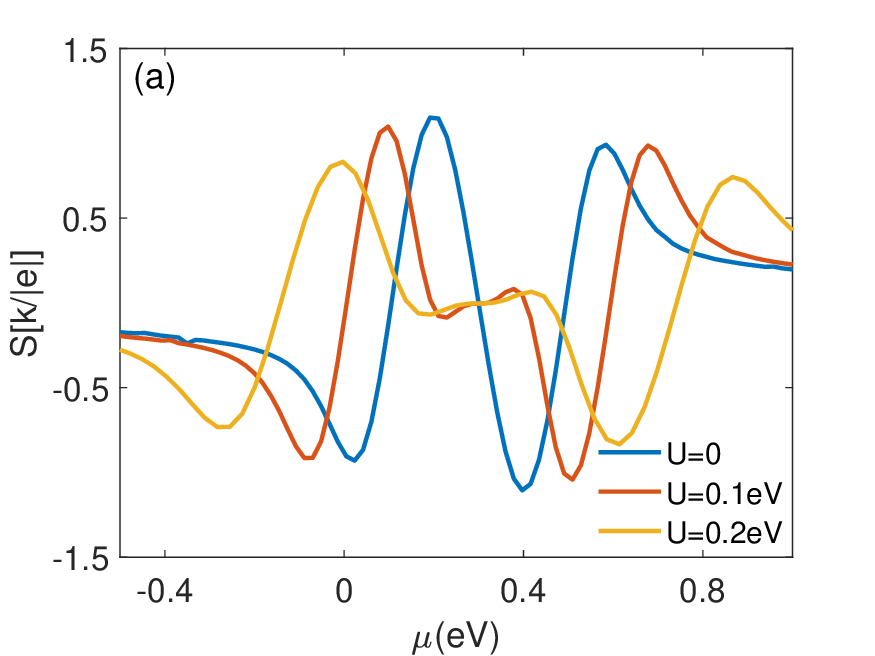}
\includegraphics[width=7.4cm,height=5.7cm]{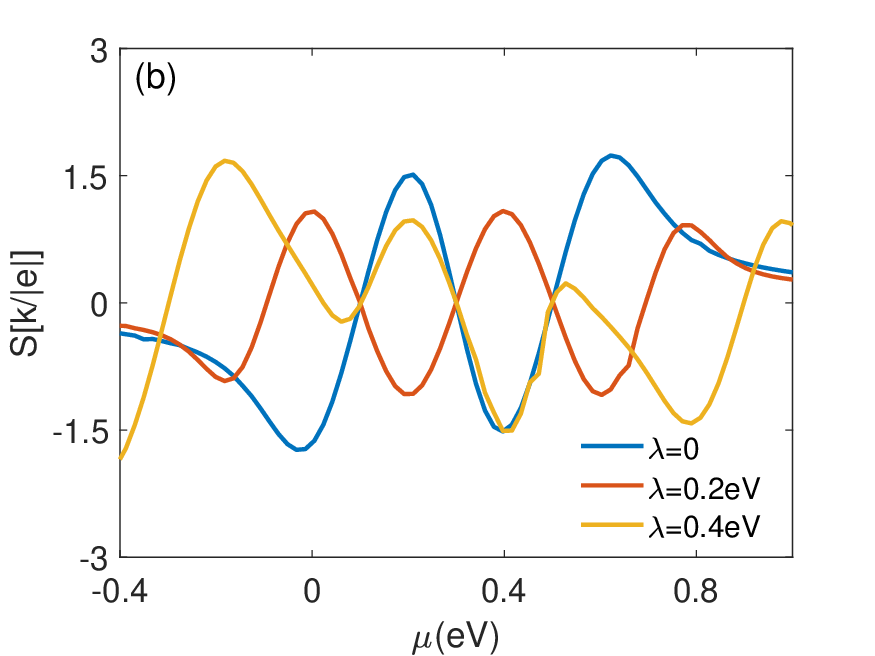}
\caption{The thermopower $ S $ as a function of the chemical potential $ \mu $ in an unbiased Marcus SMJ computed at $ kT=0.0126 eV,\, \Gamma=0.03 eV,\, \epsilon_0=0.3eV,\,  \lambda=0.2 eV$(a) and $ U=0.4 eV$(b). 
}
\label{rateI}
\end{center}\end{figure} 

\begin{figure}[t] 
\begin{center}
\includegraphics[width=7.4cm,height=5.5cm]{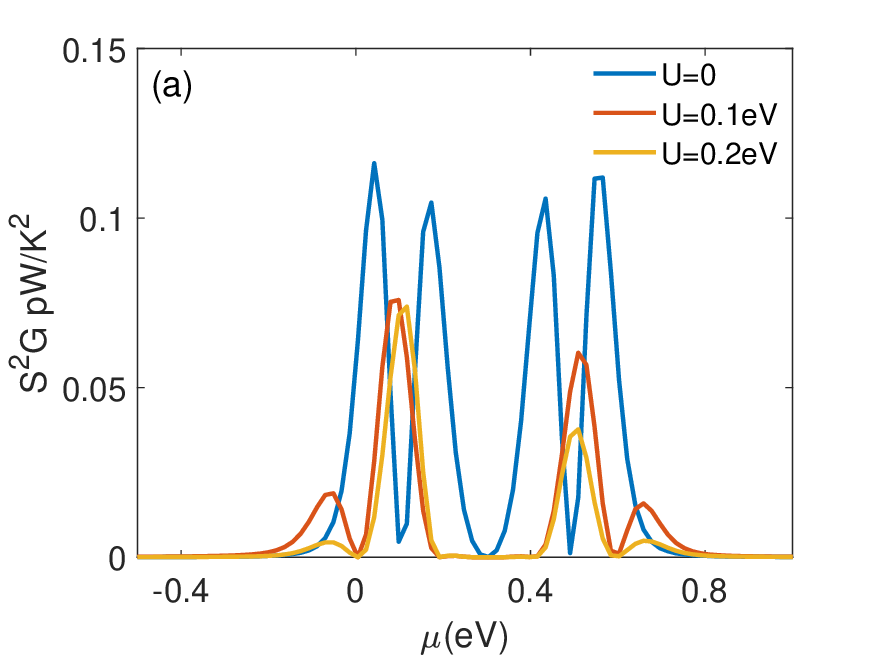}
\includegraphics[width=7.4cm,height=5.5cm]{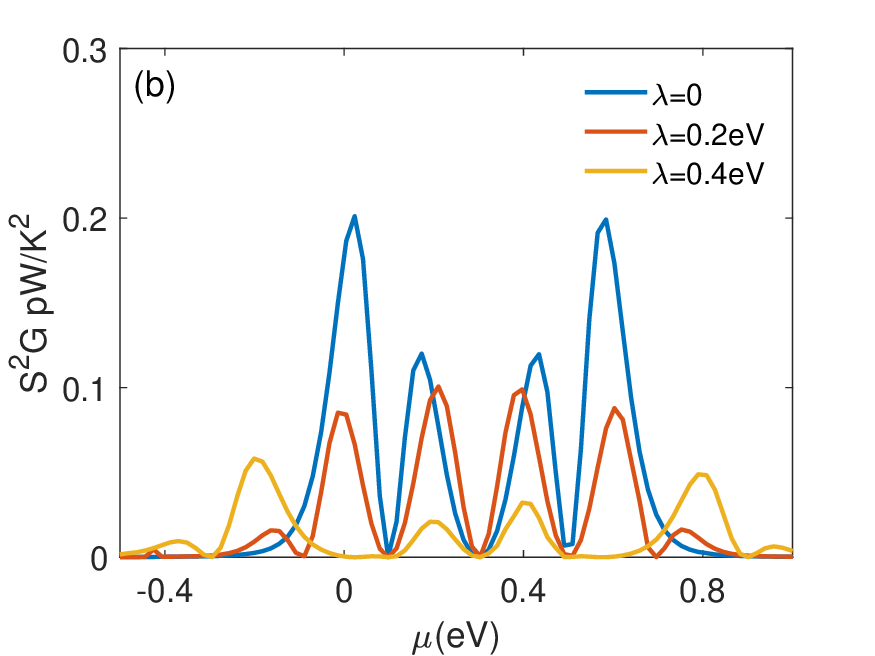}
\caption{Power factor $ S^2G $ as a function of the chemical potential of electrodes $ \mu $ computed at zero bias voltage
computed at $T_L > T_R\, , kT = 0.0126 eV,\, \Gamma = 0.03 eV,\, \epsilon_0 = 0.3 eV, \,  \lambda = 0.2 eV$(a) and $ U = 0.4 eV$(b).} 
\label{rateI}
\end{center}\end{figure}

Note that Coulomb interactions do not significantly change the conduction
values at a fixed reorganization energy. On the contrary, varying $ \lambda $ one could dramatically change 
the magnitude of conductance. This is illustrated in Fig. 2b where the curves  are plotted at a fixed charge constant $ U $ and several values of the reorganization energy. It is shown that when solvent reorganization accompanying the charge transport intensifies the conduction $G $ swiftly falls. 
        Insets displayed in Fig. 2 help to better demonstrate the changes in the molecular bridge  states occurring due to the Coulomb interactions. The inset put at Fig 2a shows the occupation probabilities for the case when Coulomb interactions are disregarded and the molecule is simulated by a two states system with a neutral state $ |b>$ and doubly charged state 
$|\tilde a> $ with the occupational probabilities equal to $P_b $ and $ 2P_a. $ In Fig. 2b the inset illustrates  the effect of Coulomb interactions on the molecular states. Here, as before, $ P_b $ is associated with the neutral
state, $2P_a = P_a^+ + P_a^-$ where $P_a^\pm $ correspond to the couple of the singly charged states and $P_c $ is associated with the double charged state.

In general, the thermopower $ S $ is known to take on nonzero values when the chemical potential of electrodes is sufficiently close to molecular level with a certain energy $ \tilde \epsilon. $ At $ \mu < \tilde \epsilon $ the corresponding molecular state takes on a part of LUMO, whereas at $\mu > \tilde\epsilon $ this state works as HOMO. When $ \mu $ approaches $\tilde \epsilon $ from below, electrons flow from the warmer electrode to the cooler one being pushed by the temperature gradient. Assuming that $T_L > T_R $, the thermopower takes negative values. As $\mu $ exceeds $\tilde\epsilon,$ the charge carriers participating in the transport through the system are holes and the thermopower becomes positive. Thus $S$ changes its sign when $ \mu $ crosses each energy level of the considered molecule.

The combined effect of Coulomb interactions between electrons on the bridge and solvent molecules brings significant changes to the thermopower behavior which is demonstrated in Fig. 3. It is shown that, unlike the conductance, the thermopower magnitude only slightly depends on the intensity of reorganization processes. However, these processes together with the Coulomb interactions significantly change profiles of the curves shown  in this figure increasing the number of points where $ S $ changes its sign. Within the adopted model, the number of such points could
   be as large as seven (See Fig. 3b). Also, as well as in the case of zero bias conductance, the counter-balance between Coulomb interactions and  intensity of reorganization processes in the solvent may occur. In such a case the thermopower takes on values close  to zero around $ \mu = \epsilon_0, $ as shown in Fig. 3a.

The efficiency  of practical realizations of single-molecule thermoelectric devices may be characterized by the power factor $ S^2G. $ The behavior of this factor is shown in Fig. 4. It so happens that the power factor strongly depends on the processes in the solvent. When the reorganization of solvent molecules intensifies, the power factor drops. This occurs mostly due to the fall of the conductance for the thermopower is significantly less sensitive to variations of the reorganization energy. Similar conclusion was made in the earlier work \cite{39}. Thus the most obvious way to enhance $ S^2G $ is to enhance the conductance by minimizing the intensity of electron-phonon coupling. Another way to increase $ G $ (and, consequently, the power factor) is to take as a bridge in the SMJ a molecule whose neutral state 
$ |b> $ is a nondegenerate one while the charge states are several-fold degenerate \cite{39}. However, we cannot consider such linkers within the adopted model.

Coulomb interactions also change the maximum values of the power factor. Besides, $ S^2 G$ turns zero around $ \mu = \epsilon_0 .$ This happens due to opposing effects of electron-electron interactions and the reorganization of solvent molecules which  cause reducing of the thermopower magnitude illustrated in Fig. 3a.

\section
		{ IV. Conclusions:}
In this work we study the effect of Coulomb interactions on the thermoelectric electron transport through single molecule junctions where the molecular bridge is put in contact with a dielectric solvent and the interactions between the traveling electrons and the solvent phonons are strong. 
      It appears that the Coulomb interactions cause the narrowing of the plateau in the $I - V $ curves which emerges around zero bias voltage and originates from the reorganization processes in the solvent. It may happen that the bridge electrons couplings with solvent phonons and with each other under zero bias are counter-balanced. This affects the thermoelectric characteristics, as shown in the figures 3a and 4a. Specifically, in such a case the thermopower takes on nearly zero values over a certain range around $ V=0. $

In general, electron interactions with the solvent controls the magnitudes of zero bias conductance in the presence as well as in the absence of Coulomb interactions. At the same time, the latter determines  characteristic features of the dependencies of zero bias thermoelectric characteristics  on the chemical potential of the electrodes. The number of maxima in the conduction and zeros in the thermopower which appear due to the molecular levels splitting originating from the Coulomb interactions ajoint with the levels splitting arising from of electron-phonon couplings. 
   As the effect of Coulomb interactions of the thermoelectric transport in Marcus junctions was not analyzed in detail so far. This work may contribute to further studies of this subject.

    \subsection
{\bf Declaration of competing interest:}
Authors declare that they have no competing financial interests or personal relationships which could influence the work reported in this paper.

    \subsection
{\bf Data availability statement:}
Data sharing is not applicable as no data are created in this study.

    \subsection
{\bf Acknowledgments:}
The work was supported by  United State NSF (DMR-PREM 2122102). The author thanks Prof. J. Otero, Drs. W. Serrano, X. Tao and G. M. Zimbovskiy for help in the manuscripts preparation.

\end{document}